\documentclass{CSML}

\def\dOi{12(1:9)2016}
\lmcsheading%
{\dOi}
{1--17}
{}
{}
{Aug.~\phantom08, 2015}
{Mar.~31, 2016}
{}

\ACMCCS{[{\bf Theory of computation}]: Logic---Finite Model Theory; Formal languages and automata theory---Tree languages}
\usepackage{latexsym}
\usepackage{amssymb,xspace,latexsym,epic,eepic,graphicx}
\usepackage{epsfig,amsmath}
\usepackage{hyperref}
\usepackage{verbatim}

\usepackage{amssymb,amsfonts}
\usepackage{amsmath}
\usepackage{amssymb,bm,latexsym}
\usepackage{stmaryrd}

\usepackage{amsmath,amsfonts,amssymb}

\newcommand{\boxtheorem}{\hfill $\Box$}
\newcommand{\sProof}[1]{\vspace{2mm}{\noindent\em Proof:~}#1\newline}

\renewcommand{\L}{{\mathcal L}}
\newcommand{\C}{{\frak C}}
\newcommand{\D}{{\frak D}} 
\newcommand{\F}{{\mathcal F}} 
\newcommand{\A}{{\mathcal A}}
\newcommand{\T}{{\mathcal T}}
\newcommand{\N}{{\mathcal N}} 
\newcommand{\B}{{\mathcal B}} 
\newcommand{\inv}{{\rm inv}} 
\newcommand{\FO}{{\rm FO}} 
\newcommand{\tp}{{\rm tp}} 
\newcommand{\qr}{{\rm qr}} 
\newcommand{\MSO}{{\rm MSO}}
\newcommand{\CMSO}{{\rm CMSO}}
\renewcommand{\SS}{{\frak S}} 

\newcommand{\All}{{\sf All}}

    \newcommand{\tree}{{\sf Trees}}  

\newcommand{\echild}{\prec_{\rm desc}}
\newcommand{\esib}{\prec_{\rm sb}}
\newcommand{\ch}{\echild}
\newcommand{\sbl}{\esib}
\newcommand{\e}{\varepsilon}

\newcommand{\dom}{\textrm{dom}}

\newcommand{\invar}[1]{(#1)_{{\rm inv}}}

\newcommand{\Fraisse}{{Fra\"\i ss\'e}}

\renewcommand{\phi}{\varphi}

\newcommand{\quant}{\textsf{Q}}

\newcommand{\Th}{{\rm Th}}

\newcommand{\CC}{\mathcal{C}}

\newcommand{\Un}{{\sf Un}}

\newcommand{\OMIT}[1]{}

\newtheorem{theorem}{Theorem}[section]

\newtheorem{proposition}[theorem]{Proposition}

\newcounter{example}
\renewcommand{\theexample}{\arabic{example}}
\newenvironment{example}{
        \refstepcounter{example}
        {\vspace{1ex}\noindent\bf Example
        \theexample}}{\boxtheorem\newline}

\begin{document}

\title[Order-Invariant Types and Their Applications]
{Order-Invariant Types and Their Applications}

\author[P.~Barcel{\'o}]{Pablo Barcel{\'o}\rsuper a}	
\address{{\lsuper a}Center for Semantic Web Research \& Department of Computer Science, 
University of Chile}	
\email{pbarcelo@dcc.uchile.cl}  
\thanks{{\lsuper a}Barcel\'o is funded by the Millennium Nucleus Center
for Semantic Web Research under Grant NC120004.} 

\author[L.~Libkin]{Leonid Libkin\rsuper b}	
\address{{\lsuper b}Laboratory for Foundations of Computer Science, School of
  Informatics, 
University of Edinburgh }	
\email{libkin@inf.ed.ac.uk}  

\thanks{{\lsuper b}Libkin is supported by
EPSRC grants J015377 and M025268.}



\begin{abstract}
Our goal is to show that the standard model-theoretic concept of types
can be applied in the study of order-invariant properties, i.e.,
properties definable in a logic in the presence of an auxiliary order
relation, but not actually dependent on that order relation. This is
somewhat surprising since order-invariant properties are more of a
combinatorial rather than a logical object. We provide two
applications of this notion. One is a proof, from the basic
principles, of a theorem by Courcelle stating that over trees,
order-invariant MSO properties are expressible in MSO with counting
quantifiers. The other is an analog of the Feferman-Vaught theorem for
order-invariant properties. 
\end{abstract}


\maketitle

\section{Introduction}

Invariant queries are an intriguing object that appear in the study of
the expressive power of logics over finite models. The interest in
them stems from the fact that to describe complexity classes by
logical means, one often needs an underlying linear order. For
instance, the Immerman-Vardi theorem characterizes polynomial time
properties of graphs as those expressible in least-fixpoint logic in
the presence of an order relation, cf.~\cite{FMTA,FMT}. However, the
ordering can be chosen arbitrarily: its only goal is to enable the
logic to simulate a Turing machine, which of course has the input on
its tape in {\em some} order. That is, 
one needs an ordering to express a property, but it
does not matter which order to use; any
order would do. Properties expressed in this fashion are called {\em
  order-invariant}. 

Since many results on capturing complexity classes require an ordering
that is used in an invariant fashion, the notion is of
interest. Before studying it for expressive logics like
least-fixpoint, one would want to understand its behavior in simpler
settings, like first-order logic (FO). Several attempts to do so,
however, show that the notion is much harder to deal with than it
initially appears.

To start with, it only makes sense for {\em finite} structures: over
infinite structures, a simple application of the interpolation theorem
shows that any order-invariant property can be expressed without the
order itself. But for finite structures, order-invariance does add
power. This was noticed (although not published) by Yuri Gurevich a
while ago, but by now this is a textbook result \cite{FMT}. The most
commonly used separating property is checking whether the number of
atoms of a Boolean algebra is even: it cannot be done in FO, but can
be if an arbitrary order on elements of the Boolean algebra is
added. More complex examples show that the separation continues to
hold if order is replaced by weaker devices such as the ability to
choose an element from a set \cite{Otto00} or the successor relation
\cite{Rossman07}.  

These observations led to the study of the power of order-invariant
properties in logics such as FO and monadic second-order logic
(MSO). Despite much effort, we still know relatively little about
order-invariant properties, and results that we know typically require
a very significant effort (see \cite{Schw12,Schw13} for
overviews). We do know nonetheless 
that order-invariant properties are
local, much like FO-definable properties themselves
\cite{GS00}, that over some tame structures such as words and trees,
order-invariance does not add power \cite{BS09,Niemisto}, and there
are results that extend invariance beyond order, for instance to
arithmetic predicates \cite{SS10,SS11}, or prove strong separation
results for auxiliary relations that are slightly weaker than order \cite{LW02}. For the more
powerful logic MSO we know that order-invariance on tame structures
such as trees boils down to adding counting to the logic \cite{Cou}. 

One of the reasons that the progress in understanding order-invariance
is rather slow is the {\em lack of logical tools} -- indeed, the set
of order-invariant properties is not really a logic, i.e., it is not
recursively enumerable. However, perhaps somewhat surprisingly in view
of this observation, some logical tools can be adapted to deal with
order-invariance. Showing this is our goal. We take the standard
model-theoretic concept of {\em types} (i.e., sets of formulae of a
logic or a fragment of a logic that hold in a given structure), which
play a prominent role in both classical and finite model theory
\cite{ck90,FMT}, and demonstrate their usefulness in the study of
order-invariance. 

Specifically, we do the following.

\begin{enumerate}
\item We define a notion of {\em order-invariant types} that extends
  the notion of types to the order-invariant setting and study its
  basic properties.
\item We show that, despite order-invariant properties not forming a
  logic, a logic-based notion of order-invariant types can actually be
  useful. We provide two applications:
  \begin{enumerate}
  \item First, we provide a proof, from the basic principles, of a
    result by Courcelle \cite{Cou} saying that over trees,
    order-invariant MSO properties are the ones expressible in MSO
    with counting quantifiers. This was reported in our conference paper
    \cite{BL05} but the proof was never published.
  \item Second, we prove an analog of the Feferman-Vaught theorem
    \cite{FV,Makowsky-apal} for order-invariant properties, and use it
    to extend the list of known classes where order-invariance does
    not increase the expressive power.
  \end{enumerate}
\end{enumerate}

\noindent While not claiming a breakthrough, the goal of this note is
to show that standard model-theoretic techniques are applicable in
this notoriously difficult area, and perhaps offer a new avenue of
attack on a host of unsolved problems related to order-invariance.

\paragraph{{\bf Organization.}} Basic concepts are defined in Section
\ref{prelim-sec}. In Section \ref{types-sec} we define order-invariant
types and study their basic properties. The proof of Courcelle's
theorem based on order-invariant types is given in Section
\ref{cou-sec}, and the order-invariant version of the Feferman-Vaught
theorem is given in Section \ref{fv-sec}.

\section{Preliminaries} 
\label{prelim-sec}

We now present basic background concepts, following \cite{FMT}.  We
assume familiarity with first-order logic (FO) and with its extension
with monadic second-order quantification known as monadic second-order
logic (MSO). First-order variables are denoted $x,y,z,\dots$, while
second-order variables are denoted $X,Y,Z,\dots$ We assume that
vocabularies are relational, i.e., they contain only relation and
constant symbols. All structures are assumed to be finite, and will be
denoted by letters $\A, \B, \ldots$; the domain of a structure
$\A$ will be denoted by $\dom(\A)$.

The {\em quantifier rank} $\qr(\phi)$ of a formula (FO or MSO) is the
depth of quantifier nesting in $\phi$. Up to logical equivalence,
there are only finitely many different formulae of quantifier rank $k$
(FO or MSO) for each given vocabulary.

With each structure $\A$ of vocabulary $\sigma$ we
associate its {\em rank-$k$ FO type} 
$$\tp_{\FO}^k(\A) \ = \ \{\phi\mid
\A \models\phi \text{ and }\qr(\phi) \leq k\},$$ 
where $\phi$ ranges
over $\FO$ sentences over $\sigma$. Similarly we define
$\tp_{\MSO}^k(\A)$. From the previous paragraph, 
both $\tp_{\FO}^k(\A)$ and $\tp_{\MSO}^k(\A)$ can be assume to be finite. 
There are finitely many rank-$k$ types, 
and 
for each rank-$k$ (FO or MSO) type $\tau$ there is a sentence 
 $\phi_\tau$ of quantifier rank $k$ in the logic that defines it, i.e., 
$\A \models\phi_\tau$ if and only if 
$\tp_{\FO}^k(\A)=\tau$ (or $\tp_{\MSO}^k(\A)=\tau$). 
In particular, $\phi_\tau = \bigwedge_{\phi \in \tp_{\FO}^k(\A)} \phi$. 
We thus
associate types with formulas that define them, and 
simply write $\tau$ instead of $\phi_\tau$. Every 
FO sentence of quantifier rank $k$ 
is equivalent to a disjunction of rank-$k$ FO types (and
likewise for MSO). 

\paragraph{{\bf EF games.}} 
For FO and MSO logical equivalence up to quantifier
rank $k$ can be captured using  Ehrenfeucht-\Fraisse\ (EF) games. The game
is played in two structures $\A$ and $\B$, over the same
vocabulary $\sigma$, by
two players, the {\em spoiler} and the {\em duplicator}, for 
$k\geq 0$ of {\em rounds}.
In round $i$
the spoiler selects a structure, say $\A$, and an element $a_i\in
\dom(\A)$; the duplicator responds by selecting an element $b_i$ in
the other structure, in this case
$b_i\in \dom(\B)$. The duplicator {\em wins}
if the mapping $a_i\mapsto b_i$, for $i \leq k$ defines a partial
  isomorphism between 
$\A$ and $\B$.  That is, for purely relational vocabulary this map is
  an isomorphism of substructures induced by $\{a_1,\ldots,a_k\}$ and
  $\{b_1,\ldots,b_k\}$. If constant symbols are involved, these sets
  are expanded by the interpretation of constants in $\A$ and $\B$,
  and the interpretation of each constant symbol $c$ in $\A$ must be
  mapped into the interpretation of that symbol in $\B$. 

The duplicator has a winning strategy in the $k$-round game if he wins in $k$ rounds
no matter how the spoiler
plays. It is well known that this happens if and only if $\A$ and $\B$
agree on all \FO\ sentences of quantifier rank up to $k$, and we 
write $\A \equiv^\FO_k \B$ to denote this.
In particular, $\A \equiv^\FO_k \B$ iff
$\tp_{\FO}^k(\A)=\tp_{\FO}^k(\B)$.  

An extension of the Ehrenfeucht-\Fraisse\ game also permits us to
determine whether two structures have the same MSO type. This
extension permits, in addition to the usual moves, also 
set moves, i.e., the spoiler can play a subset of a structure, say
$U_i\subseteq \dom(\A)$, and the duplicator must then respond with a
set in the other structure, i.e., $V_i\subseteq \dom(\B)$.
The winning condition is that the usual element moves form a partial
isomorphism of substructures expanded with predicates for the set
moves. In particular, if $a_i, U_j$ are an element and a set moves in
$\A$, and $b_i, V_j$ are the responses in $\B$, then $a_i\in U_j$ iff
$b_i\in V_j$.

As for \FO, 
the spoiler has a winning strategy in this $k$-round MSO game iff $\A$
and $\B$ agree on all MSO sentences of quantifier rank up to $k$,
i.e., iff $\tp_{\MSO}^k(\A)=\tp_{\MSO}^k(\B)$. In this case we write
 $\A \equiv_k^\MSO \B$.

\paragraph{{\bf $\C$-invariant sentences.}}
Let $\L$ be either FO or MSO and assume that $\sigma$ and $\sigma'$
are disjoint vocabularies. 
Consider structures $\A$ and $\B$ over $\sigma$ and $\sigma'$,
respectively, such that $\A$ and $\B$ share the same domain (i.e.,
$\dom(\A) = \dom(\B)$). 
We denote by $(\A,\B)$ the structure over $\sigma \cup
\sigma'$ whose domain coincides with that  of $\A$ and $\B$, and
the interpretation of $\sigma$ (resp. $\sigma'$) is inherited from
$\A$ (resp. $\B$). 

Assume that $\C$ and $\D$ are classes of structures over $\sigma'$ and
$\sigma$, respectively. An $\L$ sentence
$\phi$ over $\sigma \cup \sigma'$ is {\em $\C$-invariant over $\D$}, if for
each structure $\A \in \D$ and any two structures $\B_1,\B_2\in \C$
with the same domain than $\A$, the following holds:  
$$(\A,\B_1) \models \phi \ \ \Longleftrightarrow \ \ (\A,\B_2) \models
\phi.$$ We denote by $\invar{\L \cup \{\C\}}^\D$ the set of
$\C$-invariant $\L$ sentences over $\D$. We omit the
superscript when $\D$ is the class of all structures over
$\sigma$. 

A $\C$-invariant $\L$ sentence $\phi$ over $\D$ 
defines a {\em query} $Q_\phi$ which is a set of structures in $\D$ as follows: 
$$\begin{array}{rcl}
\A \in Q_\phi & 
\Leftrightarrow & (\A,\B)\models\phi\ \text{for some }\B\in\C\text{
  with }\dom(\A)=\dom(\B) \\
& \Leftrightarrow & (\A,\B)\models\phi\ \text{for all }\B\in\C\text{
  with }\dom(\A)=\dom(\B)\,. \\
\end{array}
$$

\noindent The most important case for us is when $\C$ is the class of linear
orders. We then write $<$ instead of $\C$, and denote by
$\invar{\L + \{<\}}^\D$ the set of {\em $<$-invariant} $\L$ sentences
over $\D$.
In other words, $\invar{\L + \{<\}}^\D$ consists of all $\L$ sentences $\phi$
over vocabulary $\sigma \cup \{<\}$ such that for every structure $\A
\in \D$, and any two linear orders $<_1,<_2$ interpreting $<$
over $\dom(\A)$, we have: $$(\A,<_1) \models \phi \ \
\Longleftrightarrow \ \ (\A,<_2) \models \phi.$$

We say that $<$-invariant $\L$ {\em collapses to} $\L$ over $\D$ if
for every sentence $\phi$ in $\invar{\L + \{<\}}^\D$, the query
$Q_\phi$ is definable in $\L$ over $\D$. It is known, for instance,
that over words and trees, $<$-invariant \FO\ collapses to \FO, see
\cite{BS09,Niemisto}. But sometimes invariance adds power, as the
example below demonstrates.

\begin{example} \label{exa:basic}  
A linear order can be used to define an MSO sentence
$\phi_{\text{even}}$ that checks
if the domain has even cardinality. Indeed, all one needs to do is to
check the existence of a subset $S$ of the ordering that corresponds to
even positions (i.e., $S$ consists of every other element starting from the second element of 
the ordering $<$) such that $S$ also contains its last element. These are
easily expressible in MSO, and hence this is an $\invar{\MSO + \{<\}}$
sentence, since it does not matter which linear order to use.

Using exactly the same idea, we can define a sentence $\quant_2
x\ \psi(x)$ checking if the number of elements $a$ satisfying a given
formula $\psi(x)$ is divisible by $2$, and in fact a  sentence $\quant_p
x\ \psi(x)$ checking if the number of elements satisfying 
$\psi(x)$ is divisible by $p$. These are known as {\em counting
  quantifiers}, and they will be important for us in the next sections.

Finally, assume that the vocabulary $\sigma$ is empty. Then the
$\invar{\MSO + \{<\}}$ sentence $\phi_{\text{even}}$ is not definable
in MSO alone, which shows that even over empty vocabularies,
$<$-invariant MSO does {\em not} collapse to MSO. The same property
shows that over the class of Boolean algebras, $<$-invariant FO does
not collapse to FO (since in Boolean algebras one can mimic MSO
quantification, c.f., \cite{FMT}). 
\end{example}

\section{$\C$-Invariant Types} 
\label{types-sec}

\newcommand{\htau}{\hat{\tau}} 

Let $\L$ be either FO or MSO. As before, $\sigma$ and $\sigma'$ are
disjoint vocabularies, $\C$ is a class of structures over
$\sigma'$ and $\D$ a class of structures over $\sigma$. 
With each structure $\A \in \D$ over $\sigma$ we define its {\em rank-$k$
$\C$-invariant $\L$ type over $\D$}, which we denote by
$\tp^k_{\invar{\L+\{\C\}}^\D}(\A)$, to be the set of all $\invar{\L \cup
\{\C\}}^\D$ sentences $\phi$ of quantifier rank at most $k$ such that 
$\A \in Q_\phi$. 

As in the case of rank-$k$ $\L$ types, each
rank-$k$ $\C$-invariant $\L$ type over $\D$ is definable by
an $\invar{\L \cup \{\C\}}^\D$ sentence of quantifier rank at most $k$. This is because 
the conjunction of all (finitely many, up to equivalence) $\C$-invariant $\L$ sentences
in $\tp^k_{\invar{\L+\{\C\}}^\D}(\A)$ is an $\invar{\L \cup
\{\C\}}^\D$ sentence of quantifier rank $k$ (because 
invariant sentences are closed under Boolean connectives). 
It follows that for any vocabularies $\sigma$ and $\sigma'$, and every $k$, there are
only finitely many rank-$k$ $\C$ invariant $\L$ types of $\D$, and
every rank-$k$ $\C$ invariant $\L$ sentence over $\D$ is equivalent to
a disjunction of such rank-$k$ types.

In this section, we provide a combinatorial characterization of rank-$k$ 
 invariant types which is crucial for our results. It can be described in terms of finite sequences of
transformations that either replace a pair of structures $(\A,\B)$
with another one of the same rank-$k$ type (the usual, not invariant),
or simply replacing the second component of the structure.  

More precisely, let $(\A,\B)$ and $(\A',\B')$ be two pairs of
structures sharing the domain (i.e., $\dom(\A)=\dom(\B)$ and
$\dom(\A')=\dom(\B')$) and let $\L, \C, \D$ be as above.  We write
$$(\A,\B) \, \sim^{\L,\D,\C}_k \, (\A',\B')$$ if either $\A=\A'$, or
$\tp_\L^k(\A,\B) = \tp_\L^k(\A',\B')$.  When $(\A',\B')$ is reachable
from $(\A,\B)$ by a finite sequence of $\sim^{\L,\D,\C}_k$ steps, 
we say that $(\A',\B')$
is a {\em $k$-flip} of $(\A,\B)$ under $(\L,\D,\C)$. The following simple observation establishes that 
$k$-flips preserve the $k$-invariant type of $\A$: 

\begin{lem} \label{lemma:ind} 
{\sloppy If $(\A',\B')$
is a {\em $k$-flip} of $(\A,\B)$ under $(\L,\D,\C)$, then we have
$\tp^k_{\invar{\L+\{\C\}}^\D}(\A) = \tp^k_{\invar{\L+\{\C\}}^\D}(\A')$. }
\end{lem} 

\proof
We show that if $(\A,\B) \sim^{\L,\D,\C}_k (\A',\B')$ then 
$\tp^k_{\invar{\L+\{\C\}}^\D}(\A) = \tp^k_{\invar{\L+\{\C\}}^\D}(\A')$. The lemma 
follows then by a straightforward induction on the length of the finite sequence of $\sim^{\L,\D,\C}_k$ 
steps that constitutes a $k$-flip. The case when $\A = \A'$ is trivial. Assume then that $\A \neq \A'$ but 
$\tp_\L^k(\A,\B) = \tp_\L^k(\A',\B')$. Consider an arbitrary sentence $\varphi \in \tp^k_{\invar{\L+\{\C\}}^\D}(\A)$. 
Then $\A \in Q_\varphi$, and therefore $(\A,\B) \models \varphi$. 
But since $\tp_\L^k(\A,\B) = \tp_\L^k(\A',\B')$, we conclude that $(\A',\B') \models \varphi$, which implies that $\A' \in Q_\varphi$. 
Therefore, $\tp^k_{\invar{\L+\{\C\}}^\D}(\A) \subseteq \tp^k_{\invar{\L+\{\C\}}^\D}(\A')$. The proof that 
$\tp^k_{\invar{\L+\{\C\}}^\D}(\A') \subseteq \tp^k_{\invar{\L+\{\C\}}^\D}(\A)$ is symmetric.
\qed 

The notion of $k$-flip describes the equivalence of invariant types. 
More formally, for $\A \in \D$ and $\B \in \C$ such that $\dom(\A) = \dom(\B)$ let us define: 
\begin{multline*}
\psi_{(\A,\B)}^k \ = \ \bigvee \{\tp^k_\L(\A',\B') \, \mid \, \A' \in \D, \, \B' \in \C, \, \dom(\A) = \dom(\B), \\ 
\text{ and } (\A',\B') \text{ is a {\em $k$-flip} of $(\A,\B)$ under $(\L,\D,\C)$}\}.
\end{multline*} 
We write $\psi_{(\A,\B)}$ instead of $\psi_{(\A,\B)}^k$ when $k$ is clear from the context. 
Then we can establish the following: 

\begin{proposition} \label{prop:flip} 
Let $k \geq 0$ and assume that $\A \in \D$ and $\B \in \C$ are structures such that $\dom(\A)=\dom(\B)$. Then 
for every $\A' \in \D$ and $\B' \in \C$ such that $\dom(\A') = \dom(\B')$ the following are equivalent: 
\begin{enumerate} 
\item $(\A',\B') \models \psi_{(\A,\B)}$. 
\item $(\A',\B')$ is a $k$-flip of $(\A,\B)$ under $(\L,\D,\C)$. 
\end{enumerate}

Furthermore, $\psi_{(\A,\B)}$ is an $\invar{\L \cup \{\C\}}^\D$ sentence of quantifier rank at most $k$ and 
$Q_{\psi_{(\A,\B)}}$ defines $\tp^k_{\invar{\L+\{\C\}}^\D}(\A)$. That is, for another structure
$\A'\in\D$, we have 
$\tp^k_{\invar{\L+\{\C\}}^\D}(\A) = \tp^k_{\invar{\L+\{\C\}}^\D}(\A')$ iff $\A' \in Q_{\psi_{(\A,\B)}}$. 
\end{proposition} 

\proof
Assume first that $(\A',\B') \models \psi_{(\A,\B)}$. Then $\tp_\L^k(\A',\B') = \tp_\L^k(\A'',\B'')$, for some $\A'' \in \D$ 
and $\B'' \in \C$ such that $(\A'',\B'')$ is a $k$-flip of $(\A,\B)$ under $(\L,\D,\C)$. But then $(\A',\B')$ is also a 
$k$-flip of $(\A,\B)$ under $(\L,\D,\C)$. Assume, on the other hand, that 
$(\A',\B')$ is a {\em $k$-flip} of $(\A,\B)$ under $(\L,\D,\C)$. Then $(\A',\B') \models \psi_{(\A,\B)}$ since $(\A',\B') \models \tp^k_\L(\A',\B')$.  

Clearly, $\psi_{(\A,\B)}$ is of quantifier rank at most $k$. 
We prove next that it is an $\invar{\L \cup \{\C\}}^\D$ sentence. Let $\A' \in \D$ and $\B_1,\B_2 \in \C$ such 
that $\dom(\A) = \dom(\B_1) = \dom(\B_2)$, and assume that $(\A',\B_1) \models \psi_{(\A,\B)}$. 
Then, from the previous characterization we have that 
$(\A',\B_1)$ is a $k$-flip of $(\A,\B)$ under $(\L,\D,\C)$. Therefore, $(\A',\B_2)$ is also 
 a $k$-flip of $(\A,\B)$ under $(\L,\D,\C)$, from which we conclude that $(\A',\B_2) \models \psi_{(\A,\B)}$.

Finally, we prove that $Q_{\psi_{(\A,\B)}}$ defines $\tp^k_{\invar{\L+\{\C\}}^\D}(\A)$. Let $\A'$ be a structure in $\D$. Assume first 
that $\tp^k_{\invar{\L+\{\C\}}^\D}(\A) = \tp^k_{\invar{\L+\{\C\}}^\D}(\A')$. Since $\psi_{(\A,\B)}$ is an $\invar{\L \cup \{\C\}}^\D$ sentence of quantifier rank $k$ 
and 
$(\A,\B) \models \psi_{(\A,\B)}$ for any $\B$ such that $\dom(\A) = \dom(\B)$, 
we conclude that $(\A',\B') \models \psi_{(\A,\B)}$ for any $\B'$ such that $\dom(\A') = \dom(\B')$. Therefore, $\A' \in Q_{\psi_{(\A,\B)}}$. 
Assume, on the other hand, that 
 $\A' \in Q_{\psi_{(\A,\B)}}$. Therefore, for any $\B' \in \C$ such that $\dom(\A') = \dom(\B')$ we have that $(\A',\B') \models \psi_{(\A,\B)}$. 
 But then $(\A',\B')$ is a $k$-flip of $(\A,\B)$ under $(\L,\D,C)$. We conclude that 
   $\tp^k_{\invar{\L+\{\C\}}^\D}(\A) = \tp^k_{\invar{\L+\{\C\}}^\D}(\A')$ from Lemma \ref{lemma:ind}.  
\qed

\section{Courcelle's Theorem from Invariant Types}
\label{cou-sec}

We have seen in Example \ref{exa:basic} that counting quantifiers
$\quant_p$ can be defined in order-invariant MSO. Such quantifiers
extend MSO by the following formation rule: if 
$\psi(x,\bar y)$ is a formula, then $\phi(\bar y)=\quant_p x
\ \psi(x,\bar y)$ is a formula. We have $\A\models\phi(\bar a)$ if 
$|\{a'\in\dom(\A)\mid \A\models \psi(a',\bar
a)\}|$ is divisible by $p$. MSO extended with such quantifiers for all
$p$ is referred to as {\em counting} MSO, or CMSO. What Example
\ref{exa:basic} tells us is that CMSO is definable in order-invariant
MSO. Courcelle's result from \cite{Cou} says that over trees, the two
coincide. 

Courcelle's proof was quite involved; it used graph grammars and an
algebraic approach to recognizability.  We now provide a much simpler
proof that uses tree automata techniques based on the invariant types
machinery. The simplest way to prove that over trees, MSO captures
tree automata, is to define, for each $k$, a deterministic tree
automaton that assigns each node the rank-$k$ MSO type of the subtree
rooted at it, cf.~\cite{FMT}. Our proof extends this to the
order-invariant setting: we show how to define an automaton that computes
order-invariant rank-$k$ types, and then prove that such an automaton
can be encoded in CMSO over trees.

\OMIT{
\paragraph{Words and trees} As it is customary when one wants to
define properties of words in terms of logical formalisms, we
represent finite words by means of logical structures. In particular,
a finite word (or string) $w = a_1 \cdots a_n$ over a finite alphabet
$\Sigma$ is represented as a structure $\A_w$ over vocabulary
$(<,(P_a)_{a \in \Sigma})$, such that the domain of $\A_w$ is
$\{1,\dots,n\}$, the interpretation of $<$ is the linear order over
the domain, and each $P_a$ is a predicate that contains exactly those
positions $1 \leq i \leq n$ such that $a_i = a$. Usually we do not
distinguish between a word $w$ and the structure $\A_w$ that
represents it. 
}

\paragraph{{\bf Trees and tree automata.}} We consider rooted trees with
oriented edges. To be able to take advantage of automata machinery, we
define them as unranked trees (cf.~\cite{tata}) without a sibling
ordering. More precisely, an
   {\em unranked tree domain} $D$ is a prefix-closed finite set of
  words of positive natural numbers such that 
$s \cdot i \in D$ implies 
$s \cdot j \in D$ for each $1 \leq j < i$. An {\em unranked
    tree} over a finite alphabet $\Sigma$ is a structure
$T = ( D,\ch,\,(P_a)_{a \in
  \Sigma} )$, 
where $D = \dom(T)$ is an unranked tree domain, 
$\ch$ is interpreted as the descendant relation
  (i.e., $s \ch s\cdot s'$ for each $s\cdot s' \in D$ such that $s'$ is
nonempty),   
and $P_a$ as the set of nodes labeled $a$, for
  each $a\in \Sigma$. As usual, we require the $P_a$'s to form a
  partition of $D$, i.e., each element of the domain is assigned a
  unique label in $\Sigma$.  The empty word will be denoted by $\e$;
  hence $\e\in D$ is the root of $T$. In order to avoid notation clutter, 
  throughout this section we simply write $\invar{\MSO+\{<\}}$ for 
  $\invar{\MSO+\{<\}}^\tree$, where $\tree$ is the set of
  all unranked trees. 

Courcelle's theorem \cite{Cou} says the following. 

\begin{thm} \label{theo:cou} \cite{Cou}  
A set of unranked trees is definable in 
$\invar{\MSO+\{<\}}$ iff it is definable in \CMSO.
\end{thm} 

As the first observation towards the proof, we note that
order-invariance can be replaced by sibling-order invariance.  A
sibling order on a tree is a binary relation $\sbl$ such that $s' \sbl
s''$ implies that $s'=s\cdot i$ and $s''=s \cdot j$ for some node $s$
and distinct numbers $i$ and $j$, and that on the set of all the
children of every node $s$ (i.e., $\{s\cdot i \mid s\cdot i \in D\}$),
the relation $\sbl$ is a linear order.  We denote the extension of the
unranked tree $T$ with sibling order $\sbl$ by $(T,\sbl)$, and call
$(T,\sbl)$ a {\em $\sbl$-ordered} unranked tree. We slightly abuse
notation and denote simply by $\sbl$ the class $\C$ of structures that
represent sibling-orders on unranked trees. We can restrict our
attention to sibling-order invariance due to the following. Indeed, it
is well known that a linear order can be defined from $\sbl$, and of
course vice versa, a linear order defines a sibling order. Since the
two are inter-definable, we have the following (as before, we write $\invar{\MSO+\{\sbl\}}$ for 
$\invar{\MSO+\{\sbl\}}^\tree$):

\begin{lem} \label{prop:order-vs-sbl} 
A set  of unranked trees is definable in  
$\invar{\MSO+\{<\}}$ iff it is definable in 
$\invar{\MSO+\{\sbl\}}$. 
\end{lem}   

Sibling order allows us to bring in tree automata over $\sbl$-ordered
unranked trees.  A {\em tree automaton} (TA) \cite{tata} is a tuple
$\N = (\Sigma,Q,F,\delta)$, where $\Sigma$ is a finite alphabet, $Q$
is a set of states, $F \subseteq Q$ is the set of final states, and
$\delta : Q \times \Sigma \rightarrow 2^{Q^*}$ is the transition
function such that $\delta(q,a)$ is a regular language over $Q$ for
every $q \in Q$ and $a \in \Sigma$.  A {\em run} of a TA $\N$ on a
$\sbl$-ordered unranked tree $(T,\sbl)$ with domain $D$ is a function
$\rho:D \rightarrow Q$ such that, for every element $s \in D$ labeled
$a$ with children $s_1 \sbl \dots \sbl s_n$, the word
$\rho(s_1) \dots \rho(s_n)$ is in $\delta(\rho(s),a)$ (if $s$
is a leaf labeled $a$, then the condition enforces that the empty word
belongs to $\delta(\rho(s),a)$). The $\sbl$-ordered tree is {\em
  accepted} by $\N$ if there is a run $\rho$ of $\N$ on $(T,\sbl)$
such that $\rho(\e)\in F$ (i.e., the root is in a state in $F$).  A
TA $\N$ is {\em deterministic} if for each $q,q' \in Q$ such that $q
\neq q'$ and $a \in \Sigma$, there is no word that belongs to
both $\delta(q,a)$ and $\delta(q',a)$. A set of $\sbl$-ordered
unranked trees is {\em regular} if and only if it is precisely the set
of trees accepted by a TA. It is well-known that TAs can be
determinized, that is, a set of $\sbl$-ordered unranked trees is
regular iff it is accepted by a deterministic TA.

We prove here that the sets of unranked trees that are
$\invar{\MSO+\{\sbl\}}$-definable can be
recognized by a particular class of TAs, which we call {\em
invariant}. This is done by extending 
traditional techniques used to establish connections
between MSO definability and automata recognizability over words and
trees. 

\begin{defi}
A TA $\N = (\Sigma,Q,F,\delta)$ is {\em $\sbl$-invariant} if for each state $q
\in Q$ and symbol $a \in \Sigma$, every permutation of a word in
$\delta(q,a)$ is also in $\delta(q,a)$. (Thus, a run of $\N$ on a
$\sbl$-ordered unranked tree does not depend on the actual
interpretation of $\sbl$). 
\end{defi} 

The next lemma establishes the desired connection between
$\invar{\MSO+\{\sbl\}}$-definability and $\sbl$-invariant TA
recognizability:  

\begin{lem} \label{lemma:crux} 
Let $S$ be an $\invar{\MSO+\{\sbl\}}$-definable set of unranked trees.  
There is a deterministic $\sbl$-invariant TA $\N$ for which $S$ is precisely
the set of unranked trees $T$ such that some $\sbl$-ordered extension 
$(T,\sbl)$ of $T$ is accepted by $\N$ (or, equivalently, each $\sbl$-ordered extension 
$(T,\sbl)$ of $T$ is accepted by $\N$).   
\end{lem}

The automaton, as we already explained, will be computing invariant
types of subtrees in its run. Before proving the lemma, we show how
Courcelle's theorem easily follows from it. One direction is
immediate from the observation made in Example \ref{exa:basic}: we saw
that counting quantifiers can be expressed in MSO using an order
relation. For the other direction,  we make use of a
description of regular
languages closed under permutations. 
Let $S_{k,p}$, for $k, p \geq 0$, be the semilinear set 
$\{k+np \mid n \in {\mathbb N}\}$.
For an alphabet $\Sigma=\{a_1,\ldots,a_r\}$, consider the Parikh map 
$\Pi: \Sigma^*\to{\mathbb N}^r$ where the $i$-th component of $\Pi(w)$
is the number of 
occurrences of $a_i$ in $w$. Then: 

\begin{lem}
\label{word-inv-lemma}
A regular language $L\subseteq \Sigma^*$ is closed under permutation
iff there exists a finite family $\SS$ of $r$-tuples of sets
of the form $S_{k,p}$, such that for each word $w$ over $\Sigma$ it is
the case that $w \in L$ iff for some 
$(S_1,\ldots,S_r)\in\SS$, we have $\Pi(w)\in S_1\times\ldots\times
S_r$.
\end{lem}

This can be obtained from results in \cite{GS66} and is also 
an immediate consequence of 
Pillig's normal form
\cite{Kozen02} which describes permutations of words in regular
languages. We also provide a simple direct model-theoretic proof of
this result in the appendix. 

Assume now that we have a set $S$ of trees that is
$\invar{\MSO+\{<\}}$-definable. By Lemma
\ref{prop:order-vs-sbl}, it is
$\invar{\MSO+\{\sbl\}}$-definable, and by Lemma
\ref{lemma:crux}, there is a $\sbl$-invariant TA $\N$ for which $S$ is
precisely the set of unranked trees $T$ for which there is a
sibling-order $\sbl$ over $\dom(T)$ such that $(T,\sbl)$ is accepted
by $\N$ (or, equivalently, for each sibling-order $\sbl$ over
$\dom(T)$ it is the case that $(T,\sbl)$ is accepted by $\N$). We
construct a CMSO sentence $\phi_S$ that precisely defines those
unranked trees $T$. This can be done by using standard techniques for
translating from tree automata into MSO (see, e.g., \cite{QA}). In
particular, $\phi_S$ expresses the existence of an accepting run of
$\N$ over some $\sbl$-ordered extension of $T$. That is, $\phi_S$
expresses that there is an assignment of states of $\N$ to the nodes
of $T$ that respects the transition function and assigns a final state
to the root of $T$. The only problem here is that the sentence
$\phi_S$ is defined over the unranked tree $T$, and hence there is no
$\sbl$-order available to check whether the transitions performed by
the run of $\N$ are valid. However, we know that $\N$ is
$\sbl$-invariant, and, therefore, that each transition of the form
$\delta(q,a)$ in $\N$ is described by a regular language $L$ that is
closed under permutation. From Lemma \ref{word-inv-lemma}, in order to
check whether a word $w$ belongs to $L$ we can simply check whether
$\Pi(w)$ belongs to some $r$-tuple of sets of the form $S_{k,p}$ in
$\SS$. This can clearly be defined with a CMSO formula since sets of
the form $S_{k,p}$ are semilinear. Hence, Courcelle's theorem follows.

\medskip

Thus, it remains to prove Lemma \ref{lemma:crux}: 

\proof 
Assume that $S$ is definable by sentence $\phi$ in
$\invar{\MSO+\{\sbl\}}$ over finite alphabet $\Sigma$.     
Let $k \geq 1$ be the 
  quantifier rank of $\phi$. We construct a
  deterministic $\sbl$-invariant TA $\N$ over alphabet
  $\Sigma$ such that the unique run of $\N$ on an arbitrary 
$\sbl$-ordered extension $(T,\sbl)$ 
of an unranked tree $T$ labels the
  root of $T$ with $\tp^k_{\invar{\MSO+\{\sbl\}}}(T)$.  

Let $\T$ be the set
of all $\tp^k_{\invar{\MSO+\{\sbl\}}}(T)$, for $T$ an unranked
tree over $\Sigma$. Assume that $T_1,\dots,T_p$ are unranked trees over
$\Sigma$.  We denote by $\A(T_1,\dots,T_p)$ the structure over
vocabulary $(P_\tau)_{\tau \in \T}$ whose domain is
$\{1,\dots,p\}$ and element $i$ ($1 \leq i \leq p$) belongs to
$P_\tau$ ($\tau \in \T$) iff 
$\tp^k_{\invar{\MSO+\{\sbl\}}}(T_i) = \tau$. Notice then that the
interpretation of the $P_\tau$'s defines a partition over the domain of
$\A(T_1,\dots,T_p)$. Let us also 
 denote by $a(T_1,\dots,
T_p)$ the unranked tree over $\Sigma$ that has a root labeled $a$ and 
trees $T_1,\dots,T_p$ hanging from this root. The following claim is
crucial for our construction of TA $\N$:  

\begin{clm} \label{claim-inv-trees} 
For every $a \in \Sigma$ and trees $T_1,\dots,T_p$ over
$\Sigma$, the type
\[\tp^k_{\invar{\MSO+\{\sbl\}}}(a(T_1,\dots,T_p))\] is 
 uniquely determined by $\tp^k_{\invar{\MSO+\{<\}}}(\A(T_1,\dots,T_p))$. 
\end{clm} 

\proof 
We slightly abuse notation and say that a sibling-ordered tree (resp., a word) is a $k$-flip of another 
sibling-ordered tree (resp., word), but formally mean that it is a $k$-flip under $(\MSO,\tree,\sbl)$ 
(resp., under $(\MSO,\All,<)$, where ${\sf All}$ is the set of all structures over the given vocabulary). 

Let $T_1,\dots,T_p,T_{p+1},\dots,T_r$ be 
unranked trees over alphabet
$\Sigma$. We show that 
for each $a \in \Sigma$ it is the case that:  
\begin{multline*}
\tp^k_{\invar{\MSO+\{<\}}}(\A(T_1,\dots,T_p)) = 
\tp^k_{\invar{\MSO+\{<\}}}(\A(T_{p+1},\dots,T_r)) \ \
\Longrightarrow \\ \tp^k_{\invar{\MSO+\{\sbl\}}}(a(T_1,\dots,T_p)) = 
\tp^k_{\invar{\MSO+\{\sbl\}}}(a(T_{p+1},\dots,T_r)).
\end{multline*} 

\noindent For each
$\tau \in \T$, let $T_{\tau}$ be an arbitrary
unranked tree such that 
$\tp^k_{\invar{\MSO+\{\sbl\}}}(T_{\tau}) = \tau$. 
Assume that for each $1 \leq j \leq r$, 
$\tp^k_{\invar{\MSO+\{\sbl\}}}(T_j) = \tau_j$. Then, from Proposition \ref{prop:flip}, for each $1 \leq j \leq r$ 
there exist sibling-orders $\sbl^{j,1}$ and
$\sbl^{j,2}$ on $\dom(T_{\tau_j})$ and $\dom(T_j)$, respectively, such that:  
\begin{equation} \label{eq2} 
\text{$(T_{\tau_j},\sbl^{j,1})$ is a $k$-flip of $(T_j,\sbl^{j,2})$.}
\end{equation}  

\noindent It is then
possible to conclude from \eqref{eq2} that 
there exist sibling-orders $\sbl^1$ and $\sbl^2$ such
that:   
$$\text{$(a(T_{\tau_1} \cdots T_{\tau_p}),\sbl^1)$ is a
$k$-flip of $(a(T_1 \cdots T_p),\sbl^2)$.}$$
In fact, $\sbl^1$ can be defined as $\sbl^{j,1}$ on the elements of
$a(T_{\tau_1} \cdots T_{\tau_p})$ that appear inside $T_{\tau_j}$ ($1
\leq j \leq p$) and as $(s_1 \sbl^1 \dots \sbl^1 s_p)$ on the
roots $s_1,\dots,s_p$ of $T_1,\dots,T_p$, respectively. 
Analogously, we can define $\sbl^2$ over $a(T_1 \cdots T_p)$, this time using
sibling-orders $\sbl^{j,2}$, for $1 \leq j \leq p$. Now the result follows by using the fact
that $(T_{\tau_j},\sbl^{j,1})$ is a $k$-flip of $(T_j,\sbl^{j,2})$, for each $1 \leq j \leq p$, and the fact that 
replacing in a sibling-ordered unranked tree $(T,\sbl)$ a subtree rooted at a
children of the root of $T$ with a sibling-ordered unranked tree of
its same rank-$k$ MSO type preserves the rank-$k$ MSO type of
$(T,\sbl)$. This fact can be proved using standard composition arguments for
the EF MSO game over sibling-ordered unranked trees 
(for a proof, see, e.g., \cite{QA}).   

Analogously, we can prove that 
there exist sibling-orders $\sbl^3$ and 
$\sbl^4$ such that: $$\text{$(a(T_{\tau_{p+1}} \cdots
T_{\tau_r}),\sbl^3)$ is a $k$-flip of $(a(T_{p+1} \cdots T_r),\sbl^4)$.}$$  
Therefore, from Proposition \ref{prop:flip} we have:  
\begin{enumerate}
\item 
$\tp^k_{\invar{\MSO+\{\sbl\}}}(a(T_{\tau_1},\dots,T_{\tau_p})) = 
\tp^k_{\invar{\MSO+\{\sbl\}}}(a(T_1,\dots,T_p))$. 
\item 
$\tp^k_{\invar{\MSO+\{\sbl\}}}(a(T_{\tau_{p+1}},\dots,T_{\tau_r})) = 
\tp^k_{\invar{\MSO+\{\sbl\}}}(a(T_{p+1},\dots,T_r))$.
\end{enumerate}   
Hence, in order to finish the proof of the claim it is sufficient to
prove  
that: 
\begin{multline*}
\tp^k_{\invar{\MSO+\{<\}}}(\A(T_1,\dots,T_p)) = 
\tp^k_{\invar{\MSO+\{<\}}}(\A(T_{p+1},\dots,T_r)) \ \ \Longrightarrow \\  
\tp^k_{\invar{\MSO+\{\sbl\}}}(a(T_{\tau_1},\dots,T_{\tau_p})) = 
\tp^k_{\invar{\MSO+\{\sbl\}}}(a(T_{\tau_{p+1}},\dots,T_{\tau_r})).
\end{multline*}  
{\sloppy This is what we do next. In order to simplify notation, 
we sometimes write $a(\kappa_1,\dots,\kappa_m)$\break and
$\A(\kappa_1,\dots,\kappa_m)$, for $\kappa_1,\dots,\kappa_m$ (not
necessarily distinct) elements in $\mathcal T$, instead of\break 
$a(T_{\kappa_1},\dots,T_{\kappa_m})$ and
$\A(T_{\kappa_1},\dots,T_{\kappa_m})$, respectively. } 

With each unranked tree of the form $T_\tau$, for $\tau \in \T$,
we associate an arbitrary sibling-order $\sbl^\tau$ over $\dom(T)$.
Given a linear order $<$ 
over $\{1,\dots,m\}$, we
associate with the unranked tree
$a(\kappa_1,\dots,\kappa_m)$, where each $\kappa_i$ is an
element in $\mathcal T$ ($1 \leq i \leq m$), a sibling-order
$\sbl^{<}$ over
$\dom(a(\kappa_1,\dots,\kappa_m))$ defined in the following
way:
\begin{enumerate}
\item The interpretation of
  $\sbl^{<}$ over
  $T_{\kappa_j}$, for $1 \leq j \leq m$, corresponds to
  $\sbl^{\kappa_j}$. 
\item Over the roots $s_1,\dots,s_m$ of
  $T_{\kappa_1},\dots,T_{\kappa_m}$, respectively, it is the case that
  $s_i \sbl^{<} s_j$ iff $i <
  j$, for each $1 \leq i,j \leq m$.  
\end{enumerate}

\noindent Because of $\tp^k_{\invar{\MSO+\{<\}}}(\A(T_1,\dots,T_p)) = 
\tp^k_{\invar{\MSO+\{<\}}}(\A(T_{p+1},\dots,T_r))$, Proposition
\ref{prop:flip} implies that there exist linear orders $<_1$
 and $<_2$ over $\dom(\A(T_1,\dots,T_p))$, respectively, $\dom(\A(T_{p+1},\dots,T_r))$ such that: 
\begin{equation*} \text{$(\A(T_1,\dots,T_p),<_1)$ is a $k$-flip of
$(\A(T_{p+1},\dots,T_r),<_2)$.}\end{equation*}  
I.e., in the
undirected graph defined by the relation $\sim_k^{\MSO,\All,<}$, both
$(\A(T_1,\dots,T_p),<_1)$ and $(\A(T_{p+1},\dots,T_r),<_2)$ belong to the same
connected component as. We prove
next that this implies that $(a(\tau_1,\dots,\tau_p),\sbl^{<_1})$ belongs to the same
connected component as $(a(\tau_{p+1},\dots,\tau_r),\sbl^{<_2})$ in the
undirected graph defined by relation $\sim_k^{\MSO,\tree,\sbl}$, i.e.,
that $(a(\tau_1,\dots, \tau_p),\sbl^{<_1})$ is a $k$-flip of 
$(a(\tau_{p+1},\dots,\tau_r),\sbl^{<_2})$. From Proposition \ref{prop:flip}, this implies
that $\tp^k_{\invar{\MSO+\{\sbl\}}}(a(\tau_1,\dots,\tau_p)) = 
\tp^k_{\invar{\MSO+\{\sbl\}}}(a(\tau_{p+1},\dots,\tau_r))$. We abuse notation 
and write $\sim_k$ instead of $\sim_k^{\MSO,\All,<}$ and $\sim_k^{\MSO,\tree,\sbl}$, as 
the superscript will always be clear from the context. 
\enlargethispage{\baselineskip}

Notice that it suffices to prove the following. If
$\kappa_1,\dots,\kappa_m,\kappa_{m+1},\dots,\kappa_n$ 
are (not necessarily distinct) elements in $\T$, 
and $<_1$ and $<_2$ are linear orders over
$\dom(\A(\kappa_1,\dots,\kappa_m)) = \{1,\dots,m\}$ and 
$\dom(\A(\kappa_{m+1},\dots,\kappa_n)) = \{1,\dots,n-m\}$, 
respectively, then 
\begin{multline*}
(\A(\kappa_1,\dots,\kappa_m),<_1) \sim_k
(\A(\kappa_{m+1},\dots,\kappa_n),<_2) \ \ \Longrightarrow \\ 
(a(\kappa_1,\dots,\kappa_m),\sbl^{<_1})
\sim_k  
(a(\kappa_{m+1},\dots,\kappa_n),\sbl^{<_2}).
\end{multline*}      
Assume then that $(\A(\kappa_1,\dots,\kappa_m),<_1) \sim_k 
(\A(\kappa_{m+1},\dots,\kappa_n),<_2)$. From the definition of relation
$\sim_k$, there are two possibilities:
\begin{enumerate}
\item 
$\A(\kappa_1,\dots,\kappa_m) = \A(\kappa_{m+1},\dots,\kappa_n)$. 
and $<_2$ is a permutation of $<_1$ over $\{1,\dots,m\}$. 
Clearly then $a(\kappa_1,\dots,\kappa_m) =
a(\kappa_{m+1},\dots,\kappa_n)$, and hence: 
$$(a(\kappa_1,\dots,\kappa_m),\sbl^{<_1})
\ \sim_k  \ 
(a(\kappa_{m+1},\dots,\kappa_n),\sbl^{<_2}).$$  
\item 
$\A(\kappa_1,\dots,\kappa_m) \neq \A(\kappa_{m+1},\dots,\kappa_n)$ but 
$(\A(\kappa_1,\dots,\kappa_m),<_1) \equiv_k^\MSO\!%
(\A(\kappa_{m+1},\dots,\kappa_n),<_2)$. 
A standard composition argument for the
EF MSO game over unranked trees  
shows in this case that $(a(\kappa_1,\dots,\kappa_m),\sbl^{<_1})
$ $\equiv_k^\MSO 
(a(\kappa_{m+1},\dots,\kappa_n),\sbl^{<_2})$ (for a proof,
see, e.g., \cite{QA}). Therefore:  
$$(a(T_{\kappa_1},\dots,T_{\kappa_m}),\sbl^{<_1})
\ \sim_k \ 
(a(T_{\kappa_{m+1}},\dots,T_{\kappa_n}),\sbl^{<_2}).$$   
\end{enumerate}   
This finishes the proof of the claim. \qed

Assume that $\T' \subseteq \T$ is the set of
$\sbl$-invariant rank-$k$ MSO types of unranked trees $T$ such that $T
\in Q_\phi$. Let us define a TA $\N :=
(\Sigma,\T,\T',\delta)$, such that the word
$\tau_1 \cdots \tau_m \in \T^*$ 
belongs to $\delta(\tau,a)$
(for $\tau \in \T$ and $a \in
\Sigma$) iff  
$\tp^k_{\invar{\MSO+\{\sbl\}}}(a(T_{\tau_1},\dots, T_{\tau_p}))
= \tau$. 
Next claim shows
that $\N$ is indeed a TA: 

\begin{clm} 
For each $a \in \Sigma$ and $\tau \in {\mathcal T}$, the set
$\delta(\tau,a)$ is a regular language over $\mathcal T$.  
\end{clm}

\proof
From Claim \ref{claim-inv-trees}, 
we know that the type $\tp^k_{\invar{\MSO+\{\sbl\}}}(a(\tau_1,\dots,\tau_p))$
is determined by 
$\tp^k_{\invar{\MSO+\{<\}}}(\A(\tau_1,\dots,\tau_p))$.
Therefore, it is sufficient to construct a deterministic NFA $\N'$ over alphabet $\T$ 
that, given a word $\tau_1 \cdots \tau_p$ in
$\T^*$, the unique run of $\N'$ over $\tau_1 \cdots \tau_p$
labels position $p$ with 
$\tp^k_{\invar{\MSO+\{<\}}}(\A(\tau_1,\dots,\tau_p))$. The
states of $\N'$ are all the types  of the form
$\tp^k_{\invar{\MSO+\{<\}}}(\A(\kappa_1,\dots,\kappa_q))$,
for $\kappa_1,\dots,\kappa_q$ (not necessarily distinct) elements in
$\T$; the initial state of $\N'$ is 
$\tp^k_{\invar{\MSO+\{<\}}}(\A())$, the rank-$k$ 
$\invar{\MSO+\{<\}}$ type of the empty structure over vocabulary
$(P_{\tau})_{\tau \in \T}$; 
and
he transition relation $\delta'$ of $\N'$ 
satisfies that $\delta(\chi,\tau)$, for $\chi =
\tp^k_{\invar{\MSO+\{<\}}}(\A(\kappa_1,\dots,\kappa_q))$ and
$\tau \in \T$, is
$\tp^k_{\invar{\MSO+\{<\}}}(\A(\kappa_1,\dots,\kappa_q,\tau))$.
We prove next that
the transition function $\delta'$ is well-defined. 
\enlargethispage{2\baselineskip}

{\sloppy We  
prove, using invariant types, that for each $\tau \in \T$, 
$\tp^k_{\invar{\MSO+\{<\}}}(\A(\kappa_1,\dots,\kappa_q,\tau))$ is
uniquely determined by
$\tp^k_{\invar{\MSO+\{<\}}}(\A(\kappa_1,\dots,\kappa_q))$. 
This shows that $\delta'$ is well-defined and also that $\N'$ is
deterministic.  
Consider then (not necessarily distinct) elements
$\kappa_1,\dots,\kappa_q,$ $\kappa_{q+1},\dots,\kappa_t$ in $\T$, and
assume that}:  
$$\tp^k_{\invar{\MSO+\{<\}}}(\A(\kappa_1,\dots,\kappa_q)) \ = \ 
\tp^k_{\invar{\MSO+\{<\}}}(\A(\kappa_{q+1},\dots,\kappa_{t})).$$
Hence by Proposition
\ref{prop:flip} there exist linear orders $<_1$ and $<_2$
over $\dom(\A(\kappa_1,\dots,\kappa_q)) = \{1,\dots,q\}$ and
$\dom(\A(\kappa_{q+1},\dots,\kappa_t)) = \{1,\dots,t-q\}$, 
respectively, such that: 
$$ \text{$(\A(\kappa_1,\dots,\kappa_q),<_1)$ is a $k$-flip of
$(\A(\kappa_{q+1},\dots,\kappa_t),<_2)$.}$$ 
But then clearly
$$ \text{$(\A(\kappa_1,\dots,\kappa_q,\tau),<'_1)$ is a $k$-flip of
$(\A(\kappa_{q+1},\dots,\kappa_t,\tau),<'_2)$,}$$ 
where $<'_1$ is the linear order over $\{1,\dots,q+1\}$ such that the
restriction of $<'_1$ over $\{1,\dots,q\}$ is $<_1$ and $i <'_1 q+1$,
for each $1 \leq i \leq q$, and equivalently for $<'_2$, this time
over $\{1,\dots,t+1-q\}$.   
We conclude from Proposition
\ref{prop:flip} that 
$$\tp^k_{\invar{\MSO+\{<\}}}(\A(\kappa_1,\dots,\kappa_q,\tau)) \ = \ 
\tp^k_{\invar{\MSO+\{<\}}}(\A(\kappa_{q+1},\dots,\kappa_{t},\tau)).$$  
This concludes the proof of the claim. 
\qed

By definition, $\N$ is
deterministic. Furthermore,  
Claim \ref{claim-inv-trees} implies that $\N$ is 
invariant. In fact, if $\tau_1 \cdots \tau_m$ is a word in ${\mathcal
T}^*$ and $\pi$ is a permutation over $\{1,\dots,m\}$, then it is the
case that  
$\tp^k_{\invar{\MSO+\{<\}}}(\A(T_{\tau_1},\dots,T_{\tau_m})) = 
\tp^k_{\invar{\MSO+\{<\}}}(\A(T_{\tau_{\pi_1}},\dots,T_{\tau_{\pi_m}}))$
(since $\A(T_{\tau_1},\dots,T_{\tau_m}) \sim_k \A(T_{\tau_{\pi_1}},\dots,T_{\tau_{\pi_m}})$, and,
therefore, Claim \ref{claim-inv-trees} tells us that  
\[\tp^k_{\invar{\MSO+\{\sbl\}}}(a(T_{\tau_1},\dots,T_{\tau_m})) = 
\tp^k_{\invar{\MSO+\{<\}}}(a(T_{\tau_{\pi_1}},\dots,T_{\tau_{\pi_m}}))\]
for each $a \in \Sigma$. 

Finally, it is possible to prove by induction that 
the unique run of $\N$ on a
$\sbl$-ordered unranked tree $(T,\sbl)$ labels each node $s$ of $T$
with $\tp^k_{\invar{\MSO+\{\sbl\}}}(T_s)$, where $T_s$ is the
subtreee of $T$ rooted on $s$. 
In fact, consider an arbitrary 
$a$-labeled node $s$ of $T$ with children $s_1 \sbl \dots \sbl s_m$,
and assume that  
the unique run of $\N$ on $(T,\sbl)$ assigns states $\tau_1,\dots,\tau_m$
to $s_1,\dots,s_m$, respectively. Then by induction
hypothesis $\tp^k_{\invar{\MSO+\{\sbl\}}}(T_{s_i}) = \tau_i$,
for each $1 \leq i \leq m$. The run of $\N$ on
$(T,\sbl)$ labels $s$ with 
$\tp^k_{\invar{\MSO+\{\sbl\}}}(a(T_{\tau_1},\dots,
T_{\tau_p}))$ by definition. But notice that 
$\A(T_{\tau_1},\dots,T_{\tau_m}) = \A(T_{s_1},\dots,T_{s_m})$, which
implies that 
$\tp^k_{\invar{\MSO+\{\sbl\}}}(a(T_{\tau_1},\dots,T_{\tau_m})) = 
\tp^k_{\invar{\MSO+\{<\}}}(a(T_{s_1},\dots,T_{s_m}))$
from 
Claim \ref{claim-inv-trees}. The result now follows since $T_s =
a(T_{s_1},\dots,T_{s_m})$.  

The latter means that for
an arbitrary sibling-order $\sbl$ over $\dom(T)$ we have that 
$\N$ accepts $(T,\sbl)$  
if and only if 
$\tp^k_{\invar{\MSO+\{\sbl\}}}(T) \in {c}'$ 
if and only if $T \in Q_\phi$. This concludes our proof.  
\qed

\section{An order-invariant Feferman-Vaught theorem}
\label{fv-sec}

The classical Feferman-Vaught theorem shows how theories of complex
structures can be recovered using theories of simpler structures they
are built from \cite{FV,Makowsky-apal}. In a simple version, it says
that \FO\ theories of product $\A\times\B$ and disjoint union
$\A\sqcup\B$ are determined by theories of $\A$ and $\B$. In the case
of disjoint unions,we assume that the vocabulary is augmented with
unary predicates for the universes of structures $\A$ and $B$. We now use
order-invariant types to show that the same is true for
order-invariant \FO\ theories of structures. As a consequence, we
obtain new classes of structures where $<$-invariant \FO\ collapses to
\FO.

Let $\Th_{\invar{\FO+<}}(\A)$ be the $<$-invariant \FO\ theory of
$\A$, i.e., the set of all $<$-invariant \FO\ sentences true in $\A$,
and $\Th^k_{\invar{\FO+<}}(\A)$ be its restriction to sentences of
quantifier rank up to $k$ (note that without such a restriction, the
theory will have a sentence describing $\A$ up to isomorphism). We
then prove the following.

\begin{thm}
\label{oi-fv-thm}
Let $\A,\B$ be structures over the same vocabulary. 
Then both $\Th^k_{\invar{\FO+<}}(\A \sqcup \B)$ and 
$\Th^k_{\invar{\FO+<}}(\A \times \B)$ are uniquely determined by
$\Th^k_{\invar{\FO+<}}(\A)$ and $\Th^k_{\invar{\FO+<}}(\B)$. 
\end{thm}

\sProof{
Given structures $\A$ and $\B$, and linear orders $<^\A,<^\B$ on
$\A$ and $\B$, respectively, we define a linear order
$(<^\A \propto <^\B)$  
on $\A \times \B$ such that $(a,b)$ precedes $(a',b')$ in this order 
whenever $b <_\B b'$, or $b = b'$ and $a <_\A a'$ (i.e.,
lexicographically, starting with the second component). It follows
immediately by a straightforward EF game argument that if we have 
structures  $\A_i,\B_i$, for $i=1,2$, over the same vocabulary,  and
linear orders $<^{\A_i}$  and $<^{\B_i}$ on them such that 
$(\A_1,<^{\A_1}) \equiv_k (\A_2,<^{\A_2})$ and $(\B_1,<^{\B_1})
\equiv_k (\B_2,<^{\B_2})$, then $(\A_1 \times \B_1,(<^{\A_1} \propto
<^{\B_1})) \equiv_k (\A_2 \times \B_2,(<^{\A_2} \propto <^{\B_2}))$. 

\OMIT{
\begin{lem} \label{lemma:good-order}
Let $\A_i,\B_i$ be structures over the same vocabulary, $i \in [1,2]$,
$<^{\A_i}$ an arbitrary linear order over $\A_i$, and $<^{\B_i}$ an arbitrary
linear order over $\B_i$. Then $(\A_1 \times \B_1,(<^{\A_1} \propto
<^{\B_1})) \equiv_k (\A_2 \times \B_2,(<^{\A_2} \propto <^{\B_2}))$ whenever 
$(\A_1,<^{\A_1}) \equiv_k (\A_2,<^{\A_2})$ and $(\B_1,<^{\B_1})
\equiv_k (\B_2,<^{\B_2})$.     
\end{lem}
}

We use this observation, and the notion of a $k$-flip, to show the
following. 

\begin{lem} \label{theo:FVgames} 
Let $\A_i,\B_i$ be structures over the same vocabulary, $i=1,2$. Then:  
\begin{enumerate}
\item
$\tp^k_{\invar{\L+\{<\}}}(\A_1 \sqcup \B_1) = \tp^k_{\invar{\L+\{<\}}}(\A_2 \sqcup \B_2)$ whenever 
$\tp^k_{\invar{\L+\{<\}}}(\A_1) = \tp^k_{\invar{\L+\{<\}}}(\A_2)$ and $\tp^k_{\invar{\L+\{<\}}}(\B_1) = \tp^k_{\invar{\L+\{<\}}}(\B_2)$. 
\item
$\tp^k_{\invar{\L+\{<\}}}(\A_1 \times \B_1) = \tp^k_{\invar{\L+\{<\}}}(\A_2 \times \B_2)$ whenever 
$\tp^k_{\invar{\L+\{<\}}}(\A_1) = \tp^k_{\invar{\L+\{<\}}}(\A_2)$ and $\tp^k_{\invar{\L+\{<\}}}(\B_1) = \tp^k_{\invar{\L+\{<\}}}(\B_2)$.
\end{enumerate} 
\end{lem}

\noindent We only prove the case of the product in Lemma \ref{theo:FVgames} since the case of
the disjoint sum  is completely analogous. 
From Proposition \ref{prop:flip} we have to show that there exist linear
orders $<^{\A_1 \times \B_1}, <^{\A_2 \times \B_2}$ on $(\A_1
\times \B_1)$ and $(\A_2 \times \B_2)$, respectively, such that 
$(\A_2 \times \B_2,<^{\A_2 \times \B_2})$ is a $k$-flip of 
$(\A_1 \times \B_1,<^{\A_1 \times \B_1})$. 

\medskip

Since 
$\tp^k_{\invar{\L+\{<\}}}(\A_1) = \tp^k_{\invar{\L+\{<\}}}(\A_2)$ and $\tp^k_{\invar{\L+\{<\}}}(\B_1) = \tp^k_{\invar{\L+\{<\}}}(\B_2)$, 
we know
from Proposition \ref{prop:flip} that there exist linear orders $<^{\A_i},
<^{\B_i}$, $i \in [1,2]$, on $\A_i$ and $\B_i$, respectively, such
that $(\A_2,<^{\A_2})$ is a $k$-flip of $(\A_1,<^{\A_1})$, and
$(\B_2,<^{\B_2})$ is a $k$-flip of $(\B_1,<^{\B_1})$. 
We now prove that $(\A_2 \times \B_2,(<^{\A_2} \propto <^{\B_2}))$ is a $k$-flip
of $(\A_1 \times \B_1,(<^{\A_1} \propto <^{\B_1}))$, which, by
Proposition \ref{prop:flip}, implies the result.  

\medskip

Assume that 
$$(\A^1,<^1_\A) \sim_k (\A^2,<^2_\A) \sim_k \ \cdots \ 
\sim_k (\A^m,<^m_\A)$$ is a sequence witnessing the fact
that $(\A_2,<^{\A_2})$ is a $k$-flip of $(\A_1,<^{\A_1})$. Then
$(\A^1,<^1_\A) = (\A_1,<^{\A_1})$, and $(\A^m,<^m_\A) =
(\A_2,<^{\A_2})$. Also, assume that 
$$(\B^1,<^1_\B) \sim_k (\B^2,<^2_\B) \sim_k \ \cdots \ 
\sim_k (\B^n,<^n_\B)$$ is a sequence witnessing the fact
that $(\B_2,<^{\B_2})$ is a $k$-flip of $(\B_1,<^{\B_1})$. Then
$(\B^1,<^1_\B) = (\B_1,<^{\B_1})$, and $(\B^n,<^n_\B) =
(\B_2,<^{\B_2})$. Then in order to show $(\A_2 \times \B_2,(<^{\A_2}
\propto <^{\B_2}))$ is a $k$-flip  
of $(\A_1 \times \B_1,(<^{\A_1} \propto <^{\B_1}))$, it is enough to
show that 
\begin{multline*}
(\A^1 \times \B^1,(<^1_\A \propto <^1_\B)) \sim_k (\A^2 \times
\B^1,(<^2_\A \propto <^1_\B)) \sim_k \ \cdots \ \sim_k (\A^m
\times \B^1,(<^m_\A \propto <^1_\B)) \sim_k \\ 
(\A^m \times \B^2,(<^m_\A \propto <^2_\B)) \sim_k (\A^m \times
\B^3,(<^m_\A \propto <^3_\B)) \sim_k \ \cdots \ \sim_k (\A^m
\times \B^n,(<^m_\A \propto <^n_\B))\,.  
\end{multline*}     
Assume first that transition is from $(\A^i \times \B^1,(<^i_\A
\propto <^1_\B))$ to $(\A^{i+1} \times \B^1,(<^{i+1}_\A \propto
<^1_\B))$ for some $i < m$. We analyze two cases:
\begin{itemize}
\item
$(\A^i,<^i_\A) \equiv_k (\A^{i+1},<^{i+1}_\A)$: We conclude 
$(\A^i \times \B^1,(<^i_\A \propto <^1_\B)) \equiv_k (\A^{i+1} \times
  \B^1,(<^{i+1}_\A \propto <^1_\B))$
from the observation at the beginning of the proof
and the fact that $(\B_1,<^1_\B) \equiv_k
  (\B_1,<^1_\B)$.   
\item 
$(\A^{i+1},<^{i+1}_\A)$ is a permutation of $(\A^i,<^i_\A)$ (i.e., one
  reinterprets the order on the same structure): Clearly,
$(\A^{i+1} \times \B^1,(<^{i+1}_\A \propto <^1_\B))$ is a
  permutation of $(\A^i \times \B^1,(<^i_\A \propto <^1_\B))$.
\end{itemize}
The case when the transition is from $(\A^m \times \B^j,(<^m_\A
\propto <^j_\B))$ to $(\A^m \times \B^{j+1},(<^{m}_\A \propto 
<^{j+1}_\B))$ for some $j < n$ 
is completely analogous. This proves the lemma. 

\medskip

We now conclude the proof of the theorem. Again, we only prove it for the product. 
Let $\phi$ be a $\invar{\FO+<}$ sentence, and assume that 
$$\big(\tp^k_{\invar{\FO+<}}(\A_1),\tp^k_{\invar{\FO+<}}(\B_1)\big),
\ \dots \ 
,\big(\tp^k_{\invar{\FO+<}}(\A_m),\tp^k_{\invar{\FO+<}}(\B_m)\big)$$ is an
enumeration of all different pairs of rank-$k$ $\invar{\FO+<}$ types of
structures $\A_i,\B_i$ such that
$(\A_i \times \B_i) \models \phi$, $i \leq m$. Associate with each
sentence $\tp^k_{\invar{\FO+<}}(\A_i)$ a propositional variable 
$\alpha_i$, and with each sentence 
$\tp^k_{\invar{\FO+<}}(\B_i)$ a propositional variable $\beta_i$. 
Then it is possible to show
that the boolean function $\Phi$ defined over propositional
variables $\alpha_i,\beta_i$ in the following way 
$$\Phi(\alpha_1,\dots,\alpha_m,\beta_1,\dots,\beta_m) = 1
\ \ \Longleftrightarrow \ \ \bigvee_i (\alpha_i \wedge \beta_i) = 1\,,$$
where $\alpha_i = 1$ if and only if $\A \models \tp^k_{\invar{\FO+<}}(\A_i)$ and
$\beta_i = 1$ if and only if $\B \models \tp^k_{\invar{\FO+<}}(\B_i)$, $i \leq
m$, satisfies that 
$$\Phi(\alpha_1,\dots,\alpha_m,\beta_1,\dots,\beta_m) = 1
\ \ \Longleftrightarrow \ \ (\A \times \B) \models \phi\,.$$  

\medskip 

\noindent In fact, assume first that $(\A \times \B) \models \phi$. Then $\A =
\A_i$ and $\B = \B_i$ for some $i \leq m$, and, therefore, 
$\alpha_i \wedge \beta_i = 1$, and $\Phi = 1$. 
Assume on the other hand that $\Phi =
1$. Then for some $i \leq m$, $\alpha_i \wedge \beta_i = 1$, implying
that $\A \models \tp^k_{\invar{\FO+<}}(\A_i)$ and $\B \models
\tp^k_{\invar{\FO+<}}(\B_i)$. Hence, $\A \equiv_k^\inv \A_i$ and $\B
\equiv_k^\inv \B_i$, and from Lemma \ref{theo:FVgames}, $(\A \times
\B) \equiv_k^\inv (\A_i \times \B_i)$. But $(\A_i \times \B_i) \models
\phi$, and thus $(\A \times \B) \models
\phi$. This completes the proof.   
\boxtheorem }

\medskip

We now use Theorem \ref{oi-fv-thm} to describe classes of
structures on which $<$-invariant \FO\ collapses to \FO. Let $\CC,
\CC'$ be classes of structures. By 
$\prod(\CC,\CC')$ (respectively, $\coprod(\CC,\CC')$) we denote
classes of structures of the form $\A\times\B$  (respectively,
$\A\sqcup\B$) where $\A\in\CC$ and $\B\in\CC'$. 

\begin{cor}
\label{collapse-cor}
 Let $\CC,
\CC'$ be classes of structures on which $<$-invariant \FO\ collapses
to \FO. Then $<$-invariant \FO\ collapses to \FO\ over both
$\prod(\CC,\CC')$ and $\coprod(\CC,\CC')$. 
\end{cor}

Indeed, every $<$-invariant sentence over $\A\times\B$ (or
$\A\sqcup\B$) is given by a finite set of order-invariant types, which in
turn, by Theorem \ref{oi-fv-thm}, are given by sets of pairs of
$<$-invariant types over $\A$ and $\B$. Since these are expressible in
\FO\ by the assumption, we get that every $<$-invariant sentence over
products or disjoint unions is expressible in \FO\ too.

\medskip

For example, combining this with the results of \cite{BS09,Niemisto}
we get that $<$-invariant \FO\ collapses to \FO\ over grids (products
of successor relations) or even products of words, i.e., grids colored
in a way that is uniquely determined by coloring of its components.

One may wonder whether we can get the collapse result for arbitrarily
colored grids. While we do not know the answer, we provide an example
that indicates not only that it is hard to obtain such a result from
the Feferman-Vaught theorem, but also that the collapse of
$<$-invariant \FO\ is a very
fragile notion.

Let $\CC$ be a class of structures. By $\Un(\CC)$ we denote the class
of structures of the form $(\A,C)$, where $\A\in\CC$ and $C$ is a
subset of $\A$ (i.e., structures of $\CC$ extended with a single unary
predicate). Already such a tiny extension can destroy the collapse.

\begin{proposition}
\label{no-collapse-prop}
There is a class of structures $\CC$ such that  $<$-invariant
\FO\ collapses to \FO\ over $\CC$, but it does not collapse to
\FO\ over $\Un(\CC)$.
\end{proposition}

\sProof{
The class $\CC$ of structures we consider will have two unary relations
$V$ and $E$ partitioning the domain, and two binary relations $L$ and
$R$. The interpretation is that elements of $E$ provide names for
edges in the complete directed graph $V \times V$, and if $e$ is the
name of an edge $(x,y)$, then $L(e,x)$ and $R(e,y)$ hold (i.e., these
stand for left and right vertices of a directed edge). 
In any \FO\ sentence over such structures (even with extra
predicates), we can first make quantification relativized to $V$ and
$E$ (i.e., $\exists x \in V$, $\exists e \in E$) and then replace each
$\exists e \in E$ with $\exists e_l, e_r \in V$, and then change each
$L(e,x)$ to $x=e_l$ and each $R(e,y)$ to $y=e_r$. If we have an
order-invariant sentence, we can assume that the order is given on $V$
and extended to $E$ lexicographically, while each element of $V$ is
below each element of $E$ (since we have a complete freedom in
choosing the order). That is, each $v < e$ is replaced by {\em true} if
$v\in V$ and $e\in E$, and each $e < e'$ is 
replaced by $e_l < e_l' \vee (e_l=e_l' \wedge e_r< e_r')$. Thus, every
order-invariant sentence over $\CC$ is equivalent to an
order-invariant sentence in the language of only one unary predicate
$V$ and the order $<$, and a simple counting argument shows that such
sentences collapse to \FO\ over $V$ (see, e.g., \cite{FMT}). Hence, we
have the collapse over $\CC$. 

To show the lack of collapse over $\Un(\CC)$, we construct an
order-invariant sentence that 
\begin{enumerate}
\item checks that the new relation $C$ is a subset of $E$, and that
  $V$ is of the form $2^X$ for some set $X$, and $C$ contains precisely the
  edges of the subset relation; 
\item and on structures of such form, uses the order to check whether
  $|X|$ is even.
\end{enumerate}

The second item is done exactly as in the proof of the fact that on
Boolean algebras, with an order one can check whether the number of
atoms is even (see Example \ref{exa:basic}), and the proof that such a
sentence is not expressible in \FO\ alone is done in exactly the same
way as the original proof separating $<$-invariant \FO\ from \FO\ on
Boolean algebras, see \cite{FMT}.

To ensure that $C$ is of the right form, we must check the following, in addition to
$C\subseteq E$.
\begin{itemize}
\item The relation $C$ is reflexive, transitive, and anti-symmetric. 
\item There is a single element $v_0\in V$ such that $v_0$ is
  connected by an edge in $C$ to every other element of $V$ (it plays
  the role of the empty set).
\item We then define a set $A$ of elements $v\neq v_0$ such that there
  is no $C$-edge between $v_0$ and some $v'$, and also a $C$-edge
  between $v'$ and $v$. These play the role of atoms of the Boolean
  algebra. We also define $A(v)$ as the set of elements of $A$ so that
  there is a $C$-edge from them to $v$ (i.e., atoms under $v$).
\item For all $v\neq v'$ which are different from $v_0$, we must have
  $A(v)\neq A(v')$.
\item For all $v\neq v'$ which are different from $v_0$, we must have
  an element $v''$ so that $A(v'')=A(v)\cup A(v')$. 
\end{itemize}
It is routine to verify that all of these are expressible in \FO\ over
$V, E, L, R$, and
ensure that $C\subseteq E$ gives $V$ and $E$ the structure of a
Boolean algebra. This concludes the proof. 
\boxtheorem
}

\paragraph{{\bf Final remarks.}} The order-invariant Feferman-Vaught
theorem presented in this section  
also holds for MSO in the case of disjoint unions, but fails for
products (see, e.g., \cite{Makowsky-apal}). There are several other
relevant operations (e.g., transductions or interpretations) that
preserve the types of structures \cite{Makowsky-apal}. It is an
interesting open problem to establish if such operations continue to
preserve $<$-invariant types as well. It also appears possible to use
the technique of invariant types to look at containment of
$<$-invariant FO in MSO (as was done in \cite{BS09}), for
instance, for unbounded disjoint unions without the extra predicates
for structures. Another possibility is to give a direct proof of a
generalization of Courcelle's theorem on linear-time data complexity
of MSO over structures of bounded treewidth to $<$-invariant MSO
\cite{CF12}, using the observation that the MSO version can be proved
using Feferman-Vaught techniques \cite{Makowsky-apal}. 
 
\section*{Acknoledgement} We are grateful to the reviewers for suggesting
  helpful pointers to the literature and modifications to improve the
  readability of the paper.

\bibliographystyle{abbrv}
\bibliography{invar.bib}


\appendix

\section{A direct proof of Lemma \ref{word-inv-lemma}. }

We prove the only if direction, as the other one is immediate. 
A word $w$ over alphabet $\Sigma=\{a_1,a_2,\dots,a_r\}$ is
said to be {\em partitioned}, if it belongs to the regular language
$(a_1)^* (a_2)^* \cdots (a_r)^*$.  Notice that for every word $w$
there is a unique permutation $w^p$ of $w$ (up to isomorphism)  
that is partitioned. 

We start by proving that rank-$k$ MSO types of partitioned words can
be defined by means of finite collections of $r$-tuples of sets of the form
$S_{k,p}$. Formally, we prove: 

\medskip \noindent (*) Let $\tau$ be the rank-$k$ MSO type of a 
partitioned word over $\Sigma$. 
There exists a finite family ${\mathcal S}_\tau$ of
$r$-tuples of sets of the form $S_{k,p}$, such that for
every word $w$ over $\Sigma$ we have that 
$\tp_\MSO^k(w_p) = \tau$ iff for some
$(S_{1},\dots,S_{r}) \in {\mathcal S}_\tau$ it is the case
that $\Pi(w) \in
S_{1} \times \dots \times S_{r}$.  

\medskip We prove (*) next.  For $a \in \Sigma$ and $w$ a partitioned
word over $\Sigma$, we denote by $w_a$ the maximal subword of $w$ that
is of the form $a^*$.  By using a standard
composition argument for the MSO EF game, one can show that for any two partitioned
words $w$ and $w'$ over $\Sigma$, $$\tp_\MSO^k(w) = \tp_\MSO^k(w') \ \ \Longleftrightarrow
\ \ \tp_\MSO^k(w_{a_i}) =  \tp_\MSO^k(w'_{a_i}), \text{ for each $1 \leq i
\leq r$}.$$ 
Hence, for each rank-$k$ MSO type $\tau$ of a partitioned
word there is a finite family $\F_\tau$ of tuples of the form
$(\tau_{a_1},\dots,\tau_{a_r})$, where each $\tau_{a_i}$ is the
rank-$k$ MSO type of some word that only uses symbol $a_i$ ($1 \leq i \leq
r$), such
that for each partitioned word $w$ the rank-$k$ MSO type of $w$ is
$\tau$ iff for some $(\tau_{a_1},\dots,\tau_{a_r}) \in {\mathcal F}_\tau$ the
rank-$k$ MSO type of $w_{a_i}$ is $\tau_{a_i}$, for each $1 \leq i \leq
r$. Therefore, in order to prove (*) it is enough to show that for
each rank-$k$ MSO type of the form $\tau_a$ ($a \in \Sigma$) there
exists a set of the form $S_{k,p}$ ($k,p \geq 0$) such that for a word
$w$ of the form $a^\star$, $$\text{the rank-$k$ MSO type of $w$ is
$\tau_a$} \ \ \Longleftrightarrow \ \ |w| \in S_{k,p}.$$ This is what
we do next.

It is known (see \cite{Ladner77} and \cite{FMT} for the textbook treatment)
that for a word $w$ of the form $a^*$, the rank-$k$
MSO type of $w$ is $\tau_a$ if and only if it is accepted by the NFA
$\N \, = \, (\{a\},\Gamma,\tau_0,\{\tau_{a}\},\delta)$, where the set
of states $\Gamma$ is 
the set of all rank-$k$ MSO types of words in $a^*$; the initial state
of $\N$ is the rank-$k$ MSO type of the empty word, denoted by
$\tau_0$; the final state of $\N$ is $\tau_{a}$; and for $\tau' \in
\Gamma$ we have that $\delta(\tau',a)$ contains all the
rank-$k$ MSO types of words of the form $w' \cdot a$, for $w'$ a
word in $a^*$ with rank-$k$ MSO type $\tau'$. Clearly, 
$\N$ is a deterministic NFA since a simple composition argument for
the MSO EF game shows that the rank-$k$ MSO type of $w'
\cdot a$ is completely determined by the rank-$k$ MSO type of
$w'$. Furthermore, since the alphabet of $\N$ is unary, a simple
inspection of the transition graph of $\N$ reveals that there must
exist integers $k,p \geq 0$ such that for a word $w$ of the form
$a^*$ it is the case that $\N$ accepts $w$ iff $|w| = k + np$ for some $n \geq 0$.
This finishes the proof of (*). We now continue with the proof of the
lemma. 

Let $L$ be a regular language that is closed under permutation. From
B\"uchi's theorem, there is an 
MSO sentence $\phi_L$ over vocabulary $(<,(P_a)_{a \in
\Sigma})$  that {\em defines} $L$, i.e., for each word $w$ 
over $\Sigma$ it is the case that $w \in L$ iff $w \models \phi_L$. 
Assume that the quantifier
rank of $\phi$ is $k$, and let $\Gamma$ be the set of all rank-$k$
MSO types of the partitioned words $w$ that satisfy $\phi$
(i.e., the ones that belong to $L$). If $\Gamma$ is
empty (which implies that $L$ is also empty), 
we define ${\mathcal S}$ to be the empty set, which clearly
satisfies the statement of the lemma.  Let us assume then that $\Gamma$ is not empty.  
We claim that a word $w$ belongs to $L$ if and only there is a
rank-$k$ MSO type $\tau \in\Gamma$ and an $r$-tuple
$(S_1,\dots,S_r) \in {\mathcal S}_\tau$ such that 
$\Pi(w) \in S_1 \times \dots \times S_r$. Indeed: 
\begin{align*} 
w \in L & \ \Longleftrightarrow  \ w^p \in L 
& \llap{($L$ is closed under permutation)}
\\
& \ \Longleftrightarrow \ \text{the
rank-$k$ MSO type $\tau$ of $w^p$ is in ${}$} & \text{($\tp_\MSO^k(w^p) \in {} \Rightarrow 
w_p \models \phi$)}
 \\
& \ \Longleftrightarrow  \  
\text{$\Pi(w) \in
S_1 \times \dots \times S_r$, for some 
$(S_1,\dots,S_r) \in {\mathcal S}_\tau$.} & \llap{(from (*))}
\end{align*} 
 This finishes the proof of the lemma since $\Gamma$ is finite.
\boxtheorem

\end{document}